\documentclass[a4paper,12pt, reqno]{article}
\usepackage[latin1]{inputenc}
\usepackage{amsmath}
\usepackage{amsfonts}
\usepackage{amssymb}
\usepackage{graphicx}

\usepackage{cite}

\usepackage[usenames, dvipsnames]{color}

\def\pp{$\pm$}
\begin{document}
	\begin{center}
		
		\LARGE{Invoking Chiral Vector Leptoquark to explain LFU violation in B Decays } \\ \vspace{.5cm}
		\large{ Bhavesh Chauhan $^{a,b\dagger}$, Bharti Kindra $^{a,b\star}$} \\ \vspace{.5cm}
		
		\small{$\quad^a$ Physical Research Laboratory, Ahmedabad, India. \\
			$\quad^b$ Indian Institute of Technology, Gandhinagar, India. \\
			$\quad^\dagger$bhavesh@prl.res.in $\quad^\star$bharti@prl.res.in 
		}
	\end{center}
	
\textbf{Abstract:} LHCb has recently reported more than $2\sigma$ deviation from the Standard Model prediction in the observable $R_{J/\psi}$. We study this anomaly in the framework of a vector leptoquark  along with other lepton flavor universality violating measurements which include $R_{K^{(*)}}$, and $R_{D^{(*)}}$. We show that a chiral vector leptoquark can explain all the aforementioned anomalies consistently while also respecting other experimental constraints. \\
	
\section{Introduction}
	Measurements of the rare decays of B mesons have shown a number of interesting deviations from the Standard Model (SM) predictions, the most recent being,
	\begin{equation}
	 R_{J/\psi}=\frac{\mathcal{BR}(B_c^+\to J/\psi \tau^+\nu_{\tau})}{\mathcal{BR}(B_c^+\to J/\psi \mu^+\nu_{\mu})}
	 \end{equation}
	LHCb recently reported \cite{lhcjpsi,lhcpres} the measured value of $R_{J/\psi}$ to be $0.71\pm 0.17\pm0.18$, which is $2\sigma$ away from the SM expectation \cite{Dutta}. At quark level, these processes involve $b \to c \ell \nu$ transition. Other anomalies based on the charged current transitions are $R_D$ and $R_D^*$ and are defined as 
	\begin{equation}
	R_{D^{(*)}}=\frac{\mathcal{BR}(\bar{B}\to D^{(*)}\tau^-\bar{\nu}_{\tau})}{\mathcal{BR}(\bar{B}\to D^{(*)}\ell^-\bar{\nu}_{\ell})}
	\end{equation} 
	where the denominator is the average value for $\ell =$ e and $\mu$.  These observables have been studied by BABAR \cite{Babarrdstar1}, Belle\cite{Bellerd1,Bellerdstar1,Bellerdstar2}, and LHCb\cite{LHCbrdstar1}, and the world average shows a deviation of 2.2$\sigma$ and 3.4$\sigma$ in $R_D$ and $R_D^*$ respectively. Other observables which show deviations involve neutral current transitions $b\to s\ell^+\ell^-$ and are defined as, 
	\begin{equation}
		R_{K^{(*)}}=\frac{\mathcal{BR}(\bar{B}\to \bar{K}^{(*)}\mu^+\mu^-)}{\mathcal{BR}(\bar{B}\to \bar{K}^{(*)}e^+e^-)}.
	\end{equation}
	
	Recent measurements of $R_{K^*}$ by LHCb show $2.1-2.3\sigma$ and $2.3-2.5\sigma$ deviations in the low-$q^2$(0.045-1.1GeV$^2$) and central-$q^2$(1.1-6GeV$^2$) regions respectively \cite{LHCbrkstar}. A deviation of 2.6$\sigma$ from SM has also been reported in $R_K$. All of these deviations hint towards lepton flavor universality violation and are independent of hadronic uncertainties in the leading order \cite{Hillerrkstar,Matiasrkstar}. This has been summarised in Table \ref{tab}. \\
	
	\begin{table}[h]
		\label{rd}
		\centering
		\begin{tabular}{|c|c|c|}
			\hline
			Observable& SM prediction & Experimental Value\\ \hline
			$R_{J/\psi}$& 0.289\pp0.010 &  0.71\pp0.17\pp0.18\\
			$R_{D}$ & 0.300\pp0.008&0.403\pp0.04\pp0.024\\
			$R_{D^*}$& 0.252\pp0.003 & 0.310\pp0.015\pp0.008\\
			$R_{K}|_{q^2=[1,6]GeV^2}$&1.0004(8) & $0.745^{+0.090}_{-0.074}\pm0.036$ \\
			$R_{K^*}|_{q^2=[0.045,1.1]GeV^2}$&0.920(7) & $0.660^{+0.110}_{-0.070}\pm0.024$ \\ 
			$R_{K^*}|_{q^2=[1.1,6]GeV^2}$& 0.996(2)& $0.685^{+0.113}_{-0.069}\pm0.047$ \\
			\hline
		\end{tabular}
		\caption{ \label{tab}Current Status of some Flavor Anomalies}
	\end{table}
	
	These anomalies have been explained in variety of frameworks including LQs \cite{HillerOne, SakakiOne, var1,var2, var3, var4, var5, varsix, var7, var8, var9, var10, var11, var12, var13, var14, var15, rukmoh, var17, var18}. In \cite{HillerOne},  LQ models which can explain $R_K$ and $R_{K^*}$ anomalies at tree level exchange are discussed, while in \cite{SakakiOne} LQ models have been tested to explain the $R_D$ and $R_D^*$ anomalies. A comparison of the two works suggest that the LQ solutions that simultaneously accommodate $R_{K^{(*)}}$ and $R_{D^{(*)}}$ are scalar LQ $S_1 \sim \left( \mathbf{\bar{3}},~\mathbf{3},~1/3\right)$ and vector LQ $U_1 \sim \left( \mathbf{3},~\mathbf{1},~2/3\right)$. In this work, we also take into account the recently measured deviation in the ratio $R_{J/\psi}$ along with other constraints from B decays and explain them using $U_1$ LQ model. \\
	
\section{Leptoquark Model}
	The interactions of $U_1$ with the SM fields are given by \cite{lqreview}, 
	\begin{equation}
	\mathcal{L} \ni (g_L)_{ij} \bar{Q}_{L}^{i,a} \gamma^\mu U_{1,\mu}L_L^{j,a}  + (g_R)_{ij} \bar{d}_{R}^{i} \gamma^\mu U_{1,\mu}e_R^{j} + (g_{\bar{R}})_{ij} \bar{u}_{R}^{i} \gamma^\mu U_{1,\mu} \nu_R^{j} .
	\end{equation}
	Albeit $U_1$ is a non-chiral LQ, we will work in a limit where the right-handed couplings are negligibly small. With this approximation, the above Lagrangian is expanded in the mass basis as, 
	\begin{equation}
	\mathcal{L} \ni (g_L)_{ij} \bar{d}_{L}^{i} \gamma^\mu U_{1,\mu}e_L^{j}  + (V\cdot g_L \cdot U)_{ij} \bar{u}_{L}^{i} \gamma^\mu U_{1,\mu} \nu_L^{j} 
	\end{equation}
	where $V$ and $U$ are the Cabibbo-Kobayashi-Maskawa (CKM) and Pontecorvo-Maki-Nakagawa-Sakata (PMNS) matrices respectively. We use the following normalization
	\begin{equation}
	g_L = g_L^0 \left( \frac{M_{LQ}}{1~\text{TeV}}\right)
	\end{equation}
	for the sake of brevity. The texture of the coupling matrix is assumed to be 
	\begin{equation}
	g_L^0 = \begin{pmatrix}
	0 & 0 & 0 \\
	0 & \lambda_1 & 0 \\
	0 & \lambda_2 & \lambda_3
	\end{pmatrix}
	\end{equation}
	so that it can accommodate $b\to s\mu^+\mu^-$ and $b\to c\tau\nu_{\tau}$ transitions. \\
 
 \section{Results and Discussion}
	We have used to following constraints as well: (a) $ \mathcal{BR}( B_s \rightarrow \mu^+ \mu^-) = 2.8^{+0.7}_{-0.6} \times 10^{-9}$ \cite{bsmumu}, and (b) $\mathcal{BR}(B \rightarrow \tau \nu) = (1.14 \pm 0.27) \times 10^{-4}$ \cite{Corwin}. We also note that for interesting region of the parameter space, the LQ contribution to $(g-2)_\mu$, $\mathcal{BR}(D_q^+\to \tau \nu )$, and $\mathcal{BR}(\tau \rightarrow \mu \gamma)$ is negligibly small. We have used the form factors presented in \cite{Dutta, FF2} to estimate $R_{J/\psi}$.\\
		
	\begin{figure}[t]
		\centering
		\includegraphics[height = 7cm]{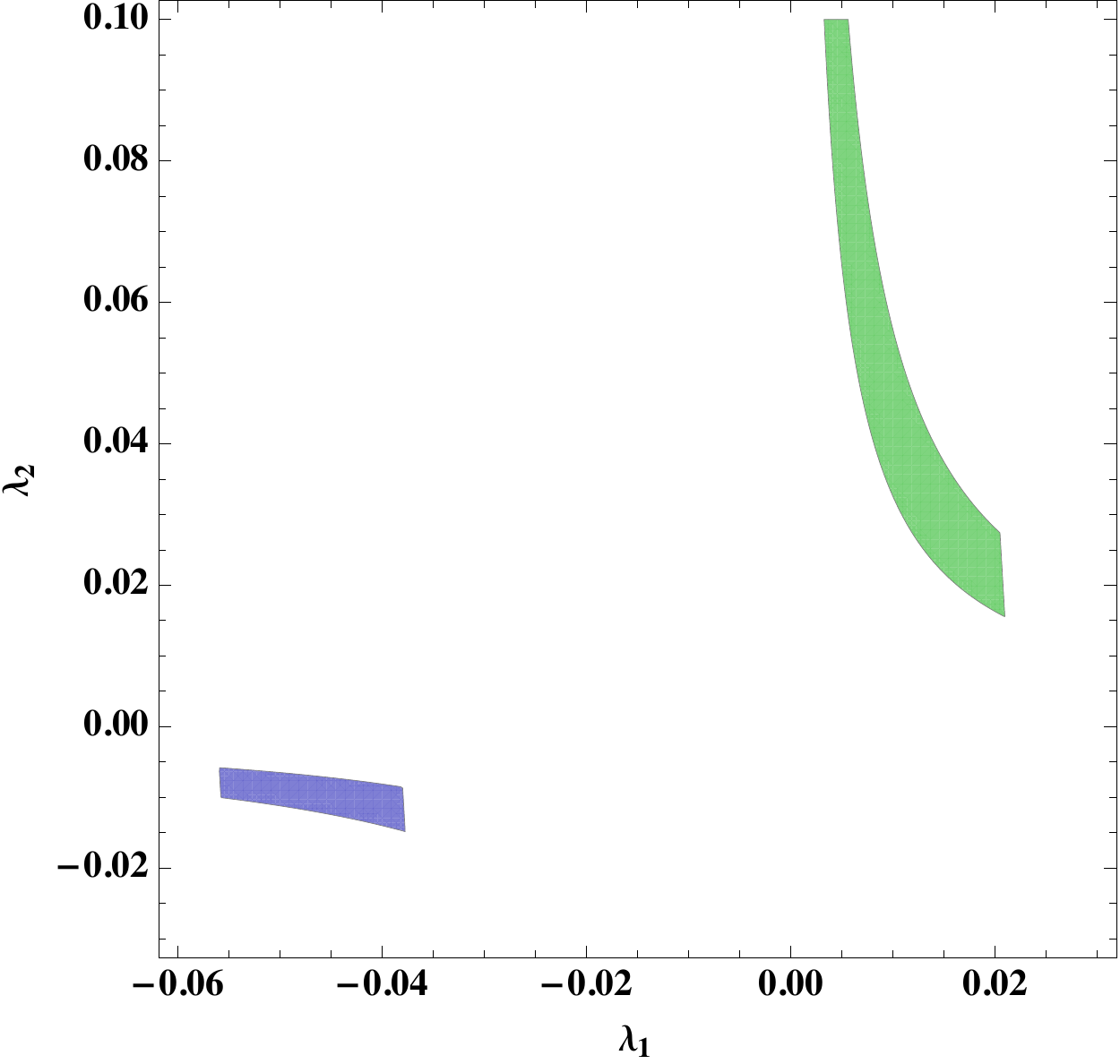}
		\caption{\label{fig1} The allowed parameter space for $\lambda_3 = 2$ and $\lambda_3 = 3$ is shown in Green and Blue respectively.}
	\end{figure}

	Since $R_{J/\psi}$ and $R_{D^*}$ are mediated by same transition of heavy to heavy quark, heavy quark spin symmetry implies that the ratios should be same at leading order \cite{LuLu}. However, the experimental values of these ratio do not seem to agree with each other in $1\sigma$ range. Because of larger uncertainties in the measured value of $R_{J/\psi}$, we take an error of $2\sigma$ in the ratio, while $R_{D^{(*)}}$ and $R_{K^{(*)}}$ are explained at $1\sigma$. \\
	
	In Fig. \ref{fig1} we show the allowed region of parameter space that explains all the mentioned flavor anomalies. \\

\section*{Acknowledgement}
The authors would like to thank Dr. Namit Mahajan for several useful discussions.

\end{document}